\documentclass{llncs}

\usepackage{graphicx}
\usepackage{textcmds} 

\begin{document}


\title{\Large A Network Protection Framework \\through Artificial Immunity}
\author{Michael Hilker \& Christoph Schommer}
\institute{University of Luxembourg\\ Faculty of Science, Technology, and Communications \\
6, Rue Richard Coudenhove-Kalergi, L-1359 Luxembourg\\
\email{\{michael.hilker, christoph.schommer\}@uni.lu}}
\maketitle

\begin{abstract}
Current network protection systems use a collection of intelligent components - e.g. classifiers or rule-based firewall systems to detect intrusions and anomalies and to secure a network against viruses, worms, or trojans. However, these network systems rely on individuality and support an architecture with less collaborative work of the protection components. They give less administration support for maintenance, but offer a large number of individual single points of failures - an ideal situation for network attacks to succeed. In this work, we discuss the required features, the performance, and the problems of a distributed protection system called {\it SANA}. It consists of a cooperative architecture, it is motivated by the human immune system, where the components correspond to artificial immune cells that are connected for their collaborative work. SANA promises a better protection against intruders than common known protection systems through an adaptive self-management while keeping the resources efficiently by an intelligent reduction of redundancies. We introduce a library of several novel and common used protection components and evaluate the performance of SANA  by a proof-of-concept implementation.
\end{abstract}


\section{Introduction}
\label{secIntroduction}
The attacks towards computer networks is increasing and the costs as well. In crime towards the computer networks does $88\%$ belong to the infections by viruses and worms but the attacks by humans is increasing. In addition, more and more attacks aim to receive information, which are afterwards sold to earn money; this thread generates a cost of about $70,000,000 \$$ \cite{CSI04}. The costs of intrusions towards computer networks is measured to $150,000,000 \$$ \cite{Sca05}. An emerging problem are the attacks performed by internal users, e.g. by unsatisfied employes of a company trying to get access to confident information. 

A network protection system is a system that tries to protect the network and its nodes against intrusions and attacks \cite{Con04}. It consists of several components of different granularity \cite{Dha05}, ranging from parameter changes in sub-components, nodes, or packet filters to powerful antivirus software, firewalls, and intrusion detection systems in important network nodes \cite{Roe99}. Each protection component contains the workflow and the knowledge in order to perform the tasks so that the protection system converges to its goals. The protection system holds the cooperation between the different protection components and the maintenance- and update workflows. The performance of the protection system is an indication how secure the network is against intrusions; the performance originates out of the performance of all protection components.

There exist different characteristics in order to evaluate and compare different protection systems \cite{Deb99}:
\begin{itemize}
\item The system has to be complete; i.e. the system protects all nodes against intrusions. 
\item The provided resources has to be used efficiently.
\item The protection system has to secure the network and all nodes. It should also not influence the user and the normal workflows in the network.
\item The installation, maintenance, and update workflows have to be easy and fast to perform. The system has to ensure that the different protection components work properly and are up-to-date.
\item The system has to adapt to the current situation in the network and it has to be adaptive in order to identify modified, mutated, and/or novel attacks.
\end{itemize}

An important characteristics of computer network protection systems is that they should be extendable and cope with upcoming - more and more complex as well as intelligent - intrusions. For example, the {\verb ZMIST } virus remains hidden in host programs \cite{Fer01}; common-used detection techniques performing pattern matching have serious problems to identify it. The {\verb EVOL } virus hides itself through swapping instructions in existing programs \cite{Szo00}. Thus, we demand for a protection system to use an adaptive information management having protection components collaborated; we demand for protection systems to be easily extendable towards and permanent protection update.

A recent approach is to model the biological systems for computer science where the artificial immune systems is one example. In this article, we introduce first a novel framework for protection systems as artificial immune system \cite{Aic04,Dec02,Hof99,Hof00}, which is motivated by the human immune system \cite{Jan04}. We then suggest a computer network protection system called SANA, present its architecture, performance and some implementation aspects. An important aspect is that SANA is a complex - but easy to understandable - system that does not use a centralised system \cite{Var02}. It provides the features and characteristics explained above and provides a dynamic, efficient, and adaptive security environment in which common used and novel approaches for network security are combined \cite{Hil06c}. These approaches are both motivated by the Biology as well as Computer Science approaches. 

\section{Current Situation}
\label{secCurrentSituation}
Current promising protection solutions and systems contain an architecture of different components. In each node, antivirus software \cite{Mik05} and firewall \cite{Ran92} are installed; they observe file access, active processes, and the network packet headers for possible attacks. Additionally, some nodes facilitate other protection components like e.g. spam filters for identifying unsolicited and undesired messages. On the network side, on hubs, switches, and routers header checking packet filters run in order to define a network policy describing which routings, protocols, and ports are allowed \cite{Mcc93}. Important nodes like e.g. internet gateways and email servers are secured using intrusion detection systems \cite{Roe99}, which check each packet completely and observe the whole node in order to identify intrusions. A novel approach are intrusion prevention systems \cite{Xin05}, which extend the antivirus software and firewall to a system observing the file access and system calls as well as checking the whole packets for intrusions. However, these systems mostly combine only antivirus and firewalls extended by some statistical approaches in order to detect unknown intrusions; however they are significantly different to intrusion detection systems.

Currently, these components must be installed and configured in each node manually where the configuration is more and more done through a centralised management server. The updates of each protection component - antivirus or intrusion detection system - works as follows: each component asks regularly a central management if there exists an update; if yes, then it downloads it. The antivirus software and the firewall ask then the user when a suspicious event occurs; the user takes over the decision and proceeds. This is often a risk since the user is mostly not an expert. Additionally, the antivirus software, firewalls, packet filters, and intrusion detection systems use a log system in order to inform the administrators about different event. The administrator analyses such, but due to the enormous number of messages is it not possible for him to analyse each message properly; the intrusions may stay undetected. 

The different protection components are not connected. Thus, each component works on its own and collaborative work does not exist. This leads primarily to redundant checks: for example, the firewall and intrusion detection system in a node check the packet for similar characteristics; resources are consequently wasted. However, more important is that the different information gathered through the different protection components are not combined in order to identify intrusions or even to identify abnormal or suspicious behaviour. Lastly, the protection system does not check itself whether each component is up-to-date and works properly, which leads to the situation that different protection components work with limited performance. An example for this problem is that the update system of an antivirus software does not work properly and e.g. the antivirus software is no longer updated and does not identify the newest intrusions; this is a problem because this node is a risk for the whole network.

For facilitating the required resources, distributed installations provide several advantages compared to the classical client server architecture. In the client server architecture, the client software - protection components - perform the tasks and demand different updates and services from the central management server. Some approaches for a distributed protection system exist \cite{Dec06}, e.g. an artificial immune system for network security \cite{Aic04,Dec02,Hof99,Hof00}. However, these systems mostly do not use a fully distributed approach, i.e. there are some centralised components that are critical for the overall performance of the system, the system is mostly not adaptive, and the system does not use small entities as artificial cells and more the multi-agent system approach \cite{Bel99} using heavy agents with few mobility, lots of internal information, and lots of knowledge about the current situation. A system is adaptive when it reconfigures itself so that it copes with the current and future situation; also learning, self-checking, and -repairing is a part of adaptive workflows. Furthermore, common used systems do not use an approach where many artificial cells have to communicate and collaborate. Furthermore, these systems are far away from production; this framework intends to narrow the gap between the distributed protection systems of academic research and production.

In the next section, a novel framework for a distributed, integrated, and dynamic protection system is introduced in which common used and novel approaches for network security are combined and which tries to solve the limitations of existing protection systems. 

\section{SANA: A Network Protection Framework through Artificial Immunity}
\label{secSANA}
\begin{figure}[p!]
\begin{center}
	\includegraphics[scale=0.7]{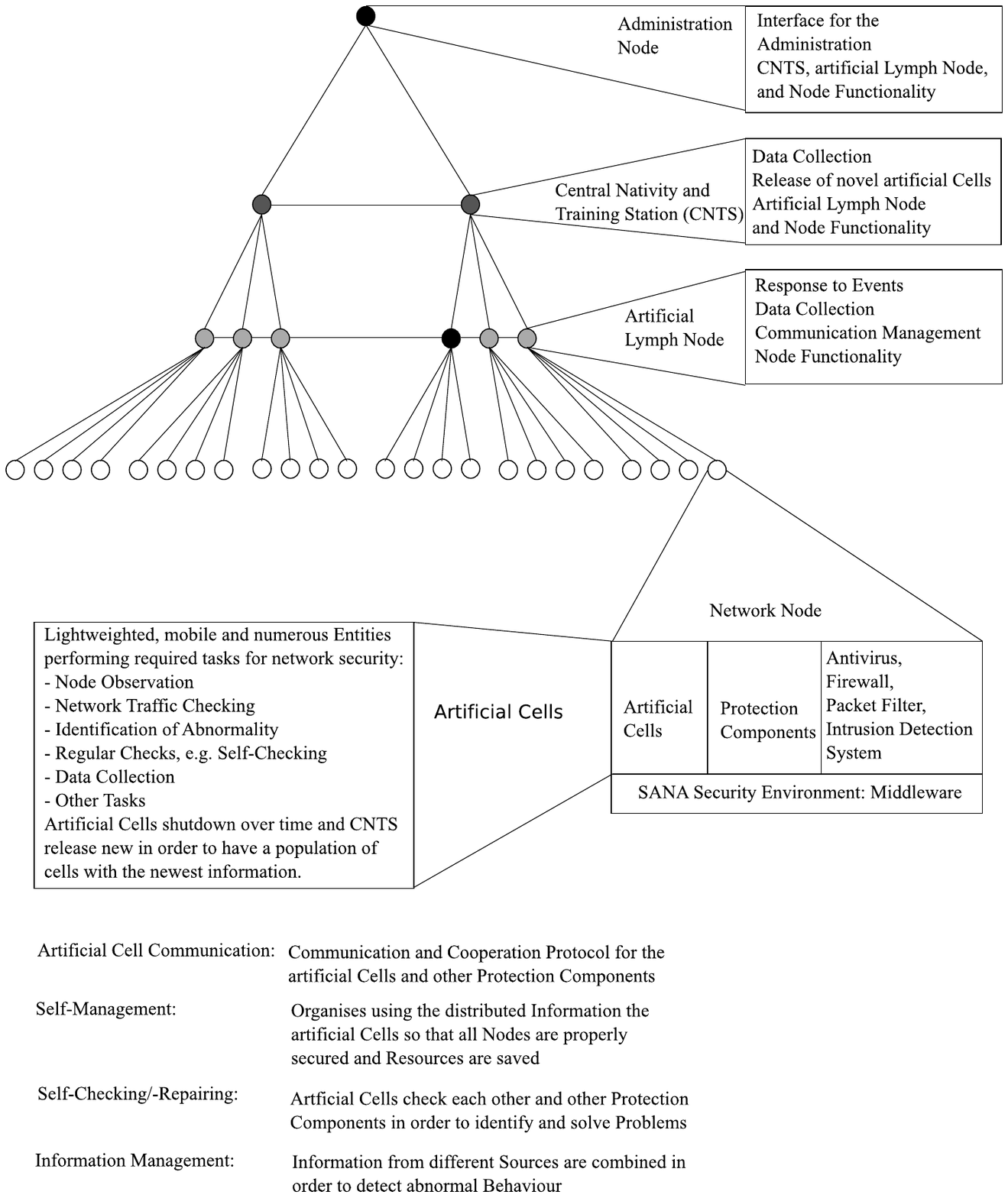}
\end{center}
\caption{Overview about SANA}
\label{pic:OverviewSana}
\end{figure}

The protection system SANA \cite{Hil06c} is a library of different network protection components using a subsymbolic organisation and information management. SANA copes with upcoming attacks that increase continuously adaption and intelligence. The administration and maintenance is simplified, it works mainly autonomously. The system does not use centralised protection components, which are single points of failure. Different components and workflows of SANA are motivated by workflows and architectures from the Biology. An overview is visualised in figure \ref{pic:OverviewSana}; the system is explained in the next sections according to this figure. 

\subsection{Security Environment}
Distributed systems with moving entities require an environment in which the entities work. Therefore, the security environment is installed in each node and provides an environment that is used by all protection components. The environment ensures the access to resources as storage, memory, CPU, and network as well as ensures that each protection component is recognised when a certain event occurs. Therefore, each protection component is installed in the security environment and registered for certain events, e.g. arriving of a packet or access to a file. Thus, the protection components are independent from the underlying hardware platform as well as network protocols and configurations, which leads to a faster deployment of protection components. The security environments is a middleware between the resources of the node and the protection components. 

The security environment also provides checks testing the protection components if they are allowed to access the resources and if they intend to perform properly. This ensures that not properly working components are removed from the network; they cannot access resources anymore. Additionally, the administrators can connect to the security environment in order to administrate these as well as to access the protection components working in this node. The administrators can quickly add novel components facilitating novel approaches of network security. 

\subsection{Protection Components}
The protection system consists of several protection components, which perform the tasks required for securing the network. These protection components are both common used protection components like antivirus software, firewall, packet filter, and intrusion detection systems in order to benefit from their performance as well as novel approaches of protection components, e.g. the artificial cells. The common used protection components are installed in the security environment and perform their tasks.\\

SANA is an artificial immune system that facilitates common protection components; SANA is an evolution of common used protection systems. The common used protection components are designed and implemented for a specific operating system. Thus, the platform independent system SANA provides an environment in order to install the protection component in the security environment. The security environment provides everything for checking, e.g. the network packets, the system calls, and the file accesses - middleware between the protection components and the resources. The protection components check these and the response is transfered to the real resources through the security environment. For other protection components, an operating systems using hardware virtualisation is provided wherein the protection components runs (see \ref{implementationsecenv}) \cite{Ope07}.

With this implementation, common used protection components as e.g. antivirus software, firewall, packet filter, and intrusion detection system are implemented and reused without internal changes in the components so that SANA is a framework for a distributed distribution system. Also, non standard approaches as e.g. virus throttle \cite{Wil02,Wil03} for slowing down the propagation of intrusions and Malfor \cite{Neu06} for automatically generating signatures of intrusions are implemented. \\

The artificial cells are dynamic, highly specialised, lightweighted, and mobile so that the system is dynamic and hard to attack. The cells are highly specialised and lightweighted because they have only limited knowledge about the situation in the network and the various tasks required for network security are distributed over several nodes. The cells are dynamic and mobile through moving around and changing of the behaviour. The number of different types of artificial cells is enormous - at least several thousands of cells - because the tasks required for network security are split in small pieces so that a breakdown of one artificial cell does not break down the whole system. Furthermore, each type of artificial cell is several times present in the network so that redundancies occur. Examples for artificial cells are introduced in \cite{Gre05,Hil06b}. Different types of artificial cells:
\begin{description}
\item[File Access] Accessing a file can install intrusions like viruses. Thus, these cells check each file that is accessed on the node. 
\item[Node Observation] Artificial cells observe the whole node including the analyse of the system calls and the control flow graphs of executed programs in order to identify and prevent installation of intrusions. 
\item[Network Traffic Analysis] This type of artificial cells evaluates each network packet whether it contains an intrusion or not. If it contains an intrusion, the artificial cells also know how to proceed with the packet, e.g. packet drop or disinfection. An example is the approach ANIMA for Intrusion Detection wherein the signatures of intrusions are stored in a network like architecture \cite{Hil06a}. 
\item[Identification of Abnormality] In the network, different examples of abnormality exist. E.g. the infected nodes or not proper working protection components must be identified and removed. AGNOSCO is an artificial cell using the information from the distributed network traffic analysis in order to identify infected nodes using the artificial ant colonies approach \cite{Hil06c}. Another approach is ANIMA for Anomaly Detection, which stores a current view of normal network traffic in order to identify and report abnormal network traffic \cite{Sch07}. 
\item[Regular Checks] Abnormality can also be recognised using regular checks. E.g. viruses and worms have a significant signature - e.g. files, running processes, and registry entries -, backdoors have an significant open port with service connected to this port, and not proper working or outdated protection components can be identified using some checks. Thus, these cells move through the network, perform the check, and identify the problems. When a problem is recognised, they inform other cells in order to solve the problem.
\item[Data Collection] These cells collect information from the network, which are used in order to monitor the network and afterwards for self-improvement and -learning.
\item[Other Types] It is possible to implement nearly all types of artificial cells where the limitations are that the cells should be small and lightweighted as well as the cells can only use the resources provided by the security environment. 
\end{description}

Using the artificial cells, it is possible to quickly introduce novel approaches; tasks as described before can be defined and performed using some artificial cells. For this, the artificial cells shutdown over time where the time is defined by the administration; values for this parameter are between some minutes and up to some hours. The systems in the network generate continuously new cells in order to keep the population up-to-date; these systems are explained in the next section. With this workflow, novel approaches and techniques for network security are quickly introduced and the newest information is included. 

\subsection{Important Nodes}
SANA contains the normal network nodes with the security environment and the protection components installed in this environment. For helping the administrators to maintain SANA, two specialised nodes are required.These are two types of nodes with more functionality, which are motivated by the human body. These specialised nodes are required in order to collect more information and to organise the system. 

\subsubsection{Artificial Lymph Nodes}
The artificial lymph nodes are a meeting point and response center for artificial cells. The artificial lymph nodes are an extension to the normal network nodes. A set of artificial lymph nodes manage a small network part - sub-sub-network called. They have additionally the following features:
\begin{itemize}
\item Collect status information about the supplied network part. 
\item Manage the communication in the sub-sub-network; see also the sectionÊ \ref{cellcomm}.
\item Respond to the messages from the artificial cells. Examples of the response are to release novel artificial cells solving a problem (disinfection a node or updating a protection component) and to inform the administrators.
\item Supply the artificial cells with additional information, e.g. whether a network packet contains an intrusion or not.
\end{itemize}

The artificial lymph nodes are redundant installed where several artificial lymph nodes manage a sub-sub-network so that a breakdown of a few interferes less the performance of the overall system. 

\subsubsection{CNTS - Central Nativity and Training Stations}
The artificial cells shutdown over time and must be replaced by new ones, which include the newest information as well as novel approaches of network security. The CNTS produce novel artificial cells and release these continuously in order to keep the number of artificial cells approximately constant. Thus, the CNTS are an extension of the artificial lymph nodes and have the same features as artificial lymph nodes as well as the generation of artificial cells. A set of CNTS manages a part of the network with several sub-sub-networks included - sub-network called. The CNTS model the bone marrow and thymus of the human immune system.

\subsection{Artificial Cell Communication} \label{cellcomm}
The artificial cells are specialised and perform only a small task. Thus, several cells have to collaborate in order to identify and prevent attacks. Therefore, the artificial cell communication is introduced, which is a distributed fault-tolerant communication protocol for point to multi-point communication. The artificial cell communication models the cell communication of the human body providing a communication and collaboration protocol for the artificial cells and other protection components. Therefore, we introduce the term {\it artificial substance}, which is used for message exchanges. It models the behaviour of different substances of the human body, e.g. the cytokines and hormones \cite{Jan04}. Each artificial substance contains the message and a header with the parameters {\verb hops-to-go } and {\verb time-to-live } describing the distribution area of the artificial substance. Each node manages the routing of the substances where each node has a set of nodes; each substance is sent to these nodes when the distribution area is not reached; this set is adapted to the current situation in the network. 

The artificial cell communication works as follows: one component sends a message, packs this message in an artificial substance, defines the distribution area, and gives it to the network. In each node, the substance is presented to all protection components and the right components receive the message. The identification of the right receivers is done using artificial receptors. These receptors are a public/private key pair describing the type and status of the artificial cells, protection components, and artificial substances. All artificial cells, protection components, and artificial substances contain several artificial receptors describing their type and status. Only when an artificial cell or protection component can authorise itself using the right keys of the artificial receptors, it will receive the message stored in the artificial substance. The protection components receiving the message respond to it, and the node sends the substance to all next nodes and the process repeats until the distribution area of the substance is reached. The next nodes are defined by administrator and adapted by the system so that the communication is configured to the current situation in the network. The artificial receptors are additionally used in order to secure the access to resources using authentication processes.

The advantages of the artificial cell communication are a distributed, efficient, and fail-safe protocol without a single point of failure. Furthermore, the protocol works fine for point to multi-point communication as used in network security where a component sends a message to all nearby components of a certain type. 

\subsection{Self Management}
The numerous artificial cells require where their tasks are most needed. Therefore, the artificial immune cells organise as follows: each protection component knows how much security it provides (security value). Each node calculates - basing on the security values of each component in the node - the security level of the node; when this level falls below a certain threshold - the threshold is defined by the administrator and defines how much security this node requires -, it starts a notification process. This notification process attracts other nearby artificial cells in order to move to this node so that the security value is increased and the artificial cells in this node are more affine to stay than to move. However, the artificial cells still work autonomously. 

For the artificial cells, performing regular tasks on the nodes, exist another workflow; we decided to use the artificial ant colonies approach during the designing of this workflow. Briefly, the artificial ants lay pheromone on the ground when carrying an artificial prey; these pheromones are afterwards used in order to find other nearby preys. Here, the preys are unchecked nodes and the artificial ants are the artificial cells. Thus, in each node $n_i$ exist a storage for values for these cells where each value $v_{n_i,n_{i+1},t_j}$ is for a connection $n_i,n_{i+1}$ and for a specific type $t_j$ of artificial cell, which is identified by the artificial receptors. When a cell performs the check in node $n_i$ and moves afterwards to node $n_{i+1}$, it increases the value for the connection $(n_i, n_{i+1})$ and for the specific cell type $t_j$ in the node $n_i$. Furthermore, the node decreases these values over time. When a cell wants to move from node $n_i$ to another node, it can use these values in order to identify the node, which waits the longest time for a check. The value describes the number of pheromones released by the artificial ants - artificial cells - where the pheromones disappear over time.

The self management increases the performance enormously because the artificial cells are properly distributed over all nodes and enough artificial cells can still keep moving in order to provide a dynamic protection system. In addition, the artificial cells performing regular tasks are lead to the nodes where this task is required; this reduces the required resources because redundant tasks are reduced and all nodes are checked regularly, which increases the security in the network.  

\subsection{Information Management}
Compared to common used protection systems, the information management of SANA is enhanced: the artificial cells analyse not only the information from one single node, they use the information from several nodes for a combined analysis. This is e.g. used for finding abnormal behaviour or multi-step/multi-stage attacks. Furthermore, the different artificial cells exchange information so that results from different analysing systems are combined evaluated and used for finding more intelligent and adaptive intrusions. Furthermore, the self-checking performed by the checking artificial cells is done distributed so that lots of information are gathered. The warnings and alerts are semi-automatically analysed and processed - common warnings and alerts are automatically processed and only novel are directly sent to the administration; They are also sent to the administrators as summary status information when required or demanded. The administrators can always demand a status report or snapshot from the system in order to receive an overview about the current situation. 

\subsection{Implementation} 
\label{implementation}
A platform-independent proof-of-concept implementation is done so that SANA can be compared with other protection systems. The simulation runs on a network simulator implementing a packet oriented network - e.g. TCP/IP - where an adversarial stresses the network and the protection system using many packets with and without intrusions. Furthermore, the implementation can be easily extended and also common used protection components as antivirus software, intrusion detection systems, and firewalls are implemented in order to evaluate and compare the performance of SANA. 

\subsubsection{Implementation of Security Environment} 
\label{implementationsecenv}
The implementation of the security environment is crucial. It should provide the features required to run the protection components efficiently but it should be impossible that the attackers use the security environment for the intrusion process.  For this, the computer is divided in four layers:
\begin{enumerate}
\item The first layer is the hardware layer containing all hardware like CPU, memory, and storage.
\item The operating system layer $1$ contains only the core parts of the operating system in order to access the resources. Mostly, these are the kernel and some drivers.
\item The operating system layer $2$ contains the rest of the operating system as well as separated the SANA security environment.
\item The software layer contains in the operating system the application software installed by the users and in the security environment the protection components. 
\end{enumerate}
\begin{figure}[t]
\begin{center}
	\includegraphics[scale=1.00]{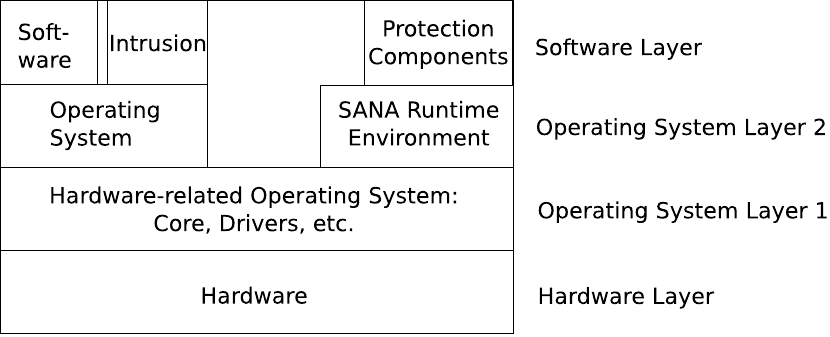}
\end{center}
\caption{Implementation layers in a network node}
\label{pic:implementationlayer}
\end{figure}

The different layers are visualised in figure \ref{pic:implementationlayer}. As it can be seen in the figure, the operating system and the SANA security environment are separated. This is achieved through hardware virtualisation. In the operating system layer $1$ is a software installed, which enables the features to run several different operating systems - guest operating systems called - at the same time. Then, one guest operating system has the operating system running and one the SANA security environment. Furthermore, the operating system cannot see the security environment but vice versa, which is ensured through the implementation. Furthermore, operating system layer $1$ is so installed that it only changes when the hardware changes. 

The advantages of this installation:\\
When an intruder installs an intrusion into the operating system or the software space, it cannot access the security environment through the hardware virtualisation. If an intruder installs an intrusion into the operating system layer $1$, the runtime environment recognises the change immediately, quarantines the node through a message to the neighbours, and removes itself from the node; then, the attacker does not receive internal information from the protection system but the administrators have to add the node to the network again. When the intruder tries to install itself in the security environment, a change is prohibited because this part does not change. Lastly, if the intruder installs an intrusive protection component - either artificial cells or local protection component - it requires the right keys from the artificial receptors in order to access the resources; without the keys, the intrusive component cannot perform intrusive tasks in order to interfere the network. Thus, the intrusion can itself only install in the software part as before and has no access to the protection system SANA. 

An important advantage is that the working operating system and the protection system runs in distinguished virtual machines. Thus, the protection system checks the operating systems and software from outside and the intrusion has more problems to hide itself from the checking because code armoring and stealth techniques often fail when the scan comes from outside. Furthermore, the virtual machines can be moved to other machines so that a user works always with its system on all machines. In addition, the protection system is implemented in a service oriented architecture (SOA) \cite{Aal06} so that more complex tasks can be demanded from the artificial lymph nodes and CNTS using different virtual machines. E.g. a not proper working security environment with protection components or a virtual machine with infected operating system can be moved to the administration computers and checked for repairing and afterwards moved back. Another example is that complex tasks are done using an additional virtual computer running the protection components performing the tasks. 

The disadvantages of the approach are an increased overhead. However, when e.g. OpenVZ is used, the overhead is about $1\%-2\%$ \cite{Ope07}. Additionally, novel CPUs already contain visualization techniques implemented in hardware so that the overhead is even more reduced. 

\subsection{Maintenance}
The maintenance is important for the administrators. The administrators demand for an easy-to-use and mostly autonomous administration and update interface where also the enhancements should be done quickly and smoothly. Therefore, SANA provides different workflows:

\subsubsection{Administration}
The administrators can connect to each security environment in order to check the current status and administrate the distributed system. The artificial lymph nodes and CNTS provide summary status information about the system and artificial cells can be released in order to collect specific information, which can be afterwards used for further analysis. 

\subsubsection{Updates}
The updates are done differently compared to common used systems where a management server provides the updates and the client softwares download it. The disadvantage is that the client software have to ask regularly for new updates and outdated software is not identified. Thus, SANA provides two different ways of updates:
\begin{enumerate}
\item The artificial cells are either updated through new information from the artificial lymph nodes, which are visited regularly. Otherwise, the most common update workflow for artificial cells is the shutdown of the cells and release of new cells with the newest updates and information.
\item The protection components are updated using updating artificial cells. These cells are released when a new update is available. They visit all protection components of this type and update these. The update is reported in order to identify unsuccessful updates or outdated components. This workflow is mostly used for not moving components. 
\end{enumerate}

In addition, the administrators can always connect to all security environments in order to access the protection components and update these.

\subsubsection{Enhancements}
Enhancements can be either released through a new population of artificial cells or an installation in the security environments. For the new population, the information about the enhancement must be added to the CNTS, which generates and releases the new artificial cells. An installation in the security environments can be either done manually through connecting to each security environment or through an automated process updating a set of security environments. With this, the enhancements are quickly deployed to the system.

\subsection{Configuration of SANA}
After the introduction of the different parts and workflows of SANA, this section provides a short overview about a possible configuration of SANA on a real network. Each network node, which are the computers in the network as well as the network equipment, contain a security environment in order to install the protection components. Each node contains an antivirus software and a firewall as in a common used protection system. The network equipment have packet filters in order to define the network policy and to close unused ports. The artificial cells move through the network and provide a distributed protection subsystem on all nodes. Important nodes, e.g. the Internet gateway and the email server but also important productive servers, run additionally to the artificial cells intrusion detection systems. The artificial lymph nodes are installed in the network equipment so that the number of hops from the nodes to the next artificial lymph nodes is small. There are two to three CNTS in the network releasing continuously novel artificial cells. The self management organises the artificial cells and guarantees a certain amount of security in each node. The artificial cell communication enables the cooperation between the different cells. All other parts are implemented as described above. 

\section{Performance and Results}
The performance of SANA is more than acceptable. It performs mostly better than common used protection systems, which are a collection of protection components, because SANA also uses the same protection components. Additionally, it contains the artificial cells providing a dynamic and adaptive part. Furthermore, SANA provides using the security environment an easy to administrate protection system that has many autonomous workflows. The warnings and alerts of each protection component are analysed and processed automatically by other protection components in order to adapt the behaviour - a more sophisticated information management. Therefore, different workflows are used: first, the information from the protection components are combined in order to identify infected nodes and abnormal behaviour; second, the infected nodes are disinfected and abnormal behaviour is observed; third, the warnings and alerts are sent to the nearby artificial cells which adapt their internal thresholds accordingly - implementation of the danger model; the danger model is an biological motivated approach where the cells exchange continuously summary status information in order to adapt their internal thresholds \cite{Aic04a,Mat02}. Furthermore, the protection system is highly dynamic: the artificial cells move continuously and provide a hard to predict and to attack protection system. Furthermore, the system can be quickly extended with novel approaches for network security. 

In the simulations, real network attacks are modelled where e.g. different worms use the network to propagate and the aim of the worm is to infect as many nodes as possible. Two simulations are performed: the first with a common used protection system consisting of antivirus software, firewall, packet filter, and intrusion detection system, and a second with SANA. In all of our simulations SANA is more secure than the common used protection system. Furthermore, SANA adapts to the current situation because identification of infected nodes, not proper working components, and suspicious behaviour is detected and the problems are fixed automatically. 

Especially, when an intrusion uses the network for propagation and an intrusion detection system checks the traffic between the network and the internet, the worm can easily infect the whole network because there is not an internal protection system that stops it. In this case, SANA'Õs artificial cells protect the nodes through distributed intrusion detection and disinfection of infected nodes. 

In the introduction, different criteria are introduced and SANA meets these:
\begin{itemize}
\item Due to the installation in all nodes SANA secures the whole network. The configuration and the self management ensures that each node is properly secured. All intrusions are not identified, which is the ideal performance of a protection system. However, SANA identifies almost all intrusions. 
\item The system uses the resources efficiently from all nodes. Thus, the required resources on a single node is reduced when SANA is used.
\item For the third criteria, the system should secure all nodes as well as it should not influence the normal production. These two criteria are met because SANA is installed on each node and uses only limited resources so that the normal production is not influenced. Furthermore due to its autonomous workflows SANA asks only rarely the user for critical security issues. 
\item The CNTS release regularly artificial cells, which check, repair, and update the different protection components for proper working - self-checking and -repairing. Furthermore, the installation is simplified because in each node must only the security environment installed and the protection components find a common infrastructure. Due to the enormous number of artificial cells is a breakdown of one cell less important for the overall performance.
\item SANA adapts to the current situation in the network. Therefore, a sophisticated information management is used and the protection components, i.e. artificial cells, identify abnormal behaviour, infected nodes, and adapt their behaviour to the current situation in the network.
\end{itemize}
After the practical implementation and analysis of the system, distributed protection systems are analysed more theoretically. Therefore, different attack scenarios are discussed where several attacks are nearly always successful when current protection systems are used. Examples are that a user wants to attack a node, shuts it therefore down, and boots from an external storage device, e.g. Linux on an USB-stick. The protection components in this system are inactive and the user can install all intrusions and other protection components of current protection system does not identify it. SANA identifies infected nodes quickly, quarantines the node, and informs the administrator. Other attacks are that novel intrusions can infect a whole network because only centralised systems identify it. In the next section, SANA is compared with common used protection systems. 

\subsection{Comparing SANA with common used Protection Systems}
As described in the section \ref{secCurrentSituation} is a common used protection system a collection of different protection components where each component is directly installed on the node. Mostly, all nodes contain an antivirus software and a firewall, all network equipments contain packet filters, as well as important nodes are secured using intrusion detection systems. In contrast, SANA facilitates the same configuration of protection components enhanced with the artificial cells, artificial lymph nodes, and CNTS. These enable collaborative, dynamic, and adaptive workflows so that SANA identifies infected nodes, weak points, and suspicious behaviour. Due to the installation of the security environment are the protection components distinguished from the operating system. Hence, the components check the system and consequently also the intrusions from outside so that it is harder for the intrusion to stealth or armour itself from the analysis. Additionally, SANA can produce a static copy of an virtual machine containing the user's operating system so that the components check this halted system. This static copy is also used to save evidences of an intrusions in order to find the adversaries, which is an important legal aspect. 

Other important network protection systems and components are distributed and host-based intrusion detection systems. Distributed intrusion detection systems install in each node an intrusion detection system or install capturing parts in each node and some centralised analysing and response centers. The second approach has the disadvantage that a lot of traffic is required in order to send all traffic to the analysing centers. The first approach installing in each node an intrusion detection system has the disadvantages that a lot of resources are required. In addition, each intrusion detection system requires lots of administration and produces several warnings and alerts, which must be analysed. Thus, the administrators cannot analyse these and, consequently, intrusions found but only reported to the log file are often not identifies. Furthermore, these approaches do not identify infected nodes, weak points in the network, and suspicious behaviour. 

Widely used in big productive and academic networks are honeypots and honeynets. A honeypot behaves like a normal computer, which pretends to contain important information and processes. Furthermore, the honeypot is normally weak secured and, thus, an aim for attacks. However, the honeypot tries to attract the intrusion in order to trap or to delay intrusions. In addition, honeypots are used in order to identify novel intrusions and generate a signature for these. Honeynets are a network of honeypots in order to simulate a whole network and to analyse the behaviour of intrusions in a network. Honeypots and honeynets are not implemented by SANA. However, it is possible to implement these protection systems in a network secured by SANA.

When a small network is used in a company and there is only few inter-network communication, the protection system is so installed that all network traffic is routed over a centralised node, which analyses all network traffic - sometimes this is implemented using a broadcast network or using adapted routing tables. Then, all other nodes must not check the network traffic and the installation of protection components is simplified. Unfortunately, this approach has the disadvantage that a lot of bandwidth is required for routing all traffic over this center as well as the center needs lots of computational power in order to check the traffic. Furthermore, when an adversarial has control over some nodes, the routing tables can be easily changed as well as the encrypted traffic can be only checked at the destination node. In addition, this configuration is static and does hardly identify suspicious behaviour and not proper working components where infected nodes can be identified through analysing the information gathered at the centralised checking node. 

The differences to current protection system are the workflows implemented through the artificial cells. With these workflows, a distributed, dynamic, and adaptive protection system is implemented, which facilitates quick and easy-to-use maintenance workflows. In addition, the information management is increased so that information from different nodes are combined analysed in order to identify and solve problems in the system. The distributed intrusion detection system deployed by the artificial cells differs SANA from common used protection systems; this distributed intrusion detection is analysed in section \ref{distrids}.

\subsection{Simulation Results}
Before different attack scenarios are discussed in detail, some simulation results comparing SANA with common used protection system are introduced. The simulations of this section focus on network based attacks where network packets containing intrusions try to infect a node. Infected nodes send packets with intrusions in order to propagate the attack. Adversaries start the attack from outside the network but infected nodes are inside. The network is always a big network with at least $1000$ nodes. 

\subsubsection{First Simulation:} 
A random network with $1110$ nodes. Three different protection systems where SANA is the implementation of SANA using only artificial cells, NIDS is the implementation of an IDS in the internet gateway, and NIDS\&SANA is combination of both. Performance: 
\begin{center}
\begin{tabular}{c|c|c|c}
Percent identified intrusive packets & NIDS & SANA & NIDS\&SANA\\\hline
First Simulation & $48.67\%$ & $79.35\%$ & $85.13\%$
\end{tabular}
\end{center}
After analysing the simulation, the results emerges out of the design. NIDS lack from the infected nodes sending lots of intrusive packets. SANA lack from the bad securing of the Internet gateway. NIDS\&SANA has the problem of infected nodes, which are not removed from the network. 

\subsubsection{Second Simulation:} 
The network is a model of a switched company network with $1110$ nodes. The configuration is similar to the first simulation. Two simulations are run: the first with $75\%$ of the traffic is external to the Internet and the second with $25\%$ external.
\begin{center}
\begin{tabular}{c|c|c|c}
Percent identified intrusive packets & NIDS & SANA & NIDS\&SANA\\\hline
Traffic $25\%$ internal & $90.64\%$ & $71.23\%$ & $94.12\%$ \\
Traffic $75\%$ internal & $37.78\%$ & $74.71\%$ & $81.74\%$
\end{tabular}
\end{center}
After analysing the simulation, the results emerges again out of the design. NIDS lack from the infected nodes sending lots of intrusive packets, which is mostly a problem in the simulation with lots of internal traffic due to no checking of this traffic. SANA lack from the bad securing of the Internet gateway where the most attacks come through. NIDS\&SANA has the problem of infected nodes, which are not identified. However, it can be seen that the combination of SANA with an IDS in the Internet gateway secures the network properly. 

\subsubsection{Third Simulation:} 
This simulation uses the same configuration as the second simulation but SANA uses the workflows to identify and disinfect infected nodes. Thus, the performance of the NIDS is the same. 
\begin{center}
\begin{tabular}{c|c|c|c}
Percent identified intrusive packets & NIDS & SANA & NIDS\&SANA\\\hline
Traffic $25\%$ internal & $90.64\%$ & $94.19\%$ & $99.20\%$ \\
Traffic $75\%$ internal & $37.78\%$ & $90.68\%$ & $95.81\%$
\end{tabular}
\end{center}
It can be seen that the infected nodes release lots of intrusive packets and, thus, the identification of infected nodes increases the performance of the protection systems enormously. 

\subsection{Performance in Attack Scenarios}
The performance of SANA is analysed through several different attack scenarios. These scenarios are very common in real life. Furthermore, most attacks can be reduced to these classes.

\subsubsection{Worm Attack}
A worm is a small software, which tries to infect a node. Therefore, it uses the network as propagation medium and exploits some security holes on the operating system and its application softwares. These security holes are mostly a badly implementation that can e.g. lead to a buffer overflow where the intrusions receive root- or superuser-rights on the node. The worm has after the infection control over the node and can collect information, erase data, and interfere the usage of the node. In addition, most worms try to propagate through sending lots of network packets containing itself or a mutated version of itself. A famous example is the \qq{ILOVEYOU} worm. 

In order to identify worm attacks, SANA facilitates different workflows. The network packets are checked through the distributed intrusion detection system with lots of artificial cells as well as through centralised intrusion detection systems in important nodes. When the worm is known by the protection system, it will remove the packets. The check is performed from outside so that armoring and stealth techniques of intrusions have more problems to hide itself. Furthermore, the information gathered through the distributed analysis is analysed in order to identify infected nodes using e.g. the artificial cells of the type AGNOSCO. Identified infected nodes are reported and checked by checking cells or by the administrators. When an infection is proved, the node is isolated and disinfected. The infection is analysed for generating signatures, which are included in the next updates. Commonly worms install e.g. backdoors, files, processes, which are characteristic for the worm. The nodes are regularly checked for different characteristics of intrusions and, when a characteristic is identified, it is reported to the administrators. 

The simulations show that SANA keeps a network clean from worm infections. Especially the identification of infected nodes and abnormal characteristics increase the performance compared to common used protection systems. The AGNOSCO approach identifies quickly infected nodes and other cells located in the nearby artificial lymph nodes start immediately a disinfection workflow. 

\subsubsection{Virus and Trojan Horse Attack}
A virus and a trojan horse attack is similar to a worm attack. The difference is that the virus requires a host program wherein it is included. When the host program is copied the virus is copied as well as when the host program is executed is the virus executed. A trojan horse is a software that primarily looks and behaves like a normal program but when it is executed it installs an intrusions. The trojan normally does not exploit a security hole, it facilitates the approach of social engineering where the user is manipulated in order to perform actions. The goal of a trojan is to install a backdoor, which is used by the adversarial in order to access and control the node. In contrast to worms and viruses does a trojan not always propagate. 

The difference between the worm and the virus and trojan horse attack are the propagation and infection workflows. The viruses and trojan horses do not always arrive using the network where mainly all data exchanges are facilitated. The nodes are observed for intrusive processes, file access, and network packets through the different protection components as e.g. artificial cells, antivirus software, and intrusion detection systems. When an infected file is identified, not the user is contacted as in common used protection systems, the systems analyses the warning autonomously and processes this. Only when the system does not know how to proceed, it contacts the administrators for further steps. The workflows to identify and disinfect infected nodes are the same as used in the worm attack scenario. 

The simulations substantiate the performance because SANA identifies and prevents known attacks as well as SANA identifies infected nodes through the tasks performed by different artificial cells. E.g. backdoors or running processes are reported and removed by the system. 

\subsubsection{Multi-Stage Attack}
Multi-stage attacks try to infect nodes or try to gather information while attacking using various steps. Each step is a sub-attack and can facilitates different attack techniques. These attacks are used in order to adapt an attack towards the protection system in the network as well as to collect information about the deployed protection system. 

The problem of these attacks is that each stage is only a small tasks, which is normally not identified as intrusive or the task is normal behaviour of programs - mimicry of normal behaviour. In the process of multiple stages, the attack can both collect information as well as find a weak point in the network in order to attack and infect the node. In order to identify the attack, the different stages must be detected and this information must be merged. Therefore, the artificial cells and other protection components observe the node and identify the different stages. This is logged and the other cells and components can read the centralised log in a node. Furthermore using the artificial cell communication, the cells exchange continuously summary status information where the appearance of these stages are included. Thus, the nearby cells - within some hops distance - receive the information about the stages and can include this into further workflows. Identification and disinfection of successful attacks is similar to the two previous attack scenarios.  

The simulations show that SANA identifies and prevents known multi-stage attacks. Unknown/novel multi-stage attacks are identified when the stages are similar to known attacks or the infection is found using the checking workflows. 

\subsubsection{Hacker Attack}
When a human attacks a network is this hacker attack called. Also the emerging class of attacks performed by humans with internal information belongs to this class. 

Attacks performed by humans are harder to identify because the humans behave in each situation different as well as unpredictable and adapt the attack to the current situation in the network. However, different tasks are always present when an intrusion is installed and SANA tries to find these characteristics. Therefore SANA observes the node as well as the network traffic in order to identify intrusive behaviour. Additionally to common used protection systems, SANA combines using the artificial cell communication and the movement of cells information from different nodes. 

The simulations show that especially backdoors e.g. installed by hackers are quickly found by checking cells as well as intrusive behaviour is identified on several nodes and reported to the administration. 

\subsubsection{Attacks using encrypted VPN Traffic}
Current networks facilitate virtual private networks (VPNs) in order to include external nodes into the system or to safely connect external services into the network. Normally, the VPNs build up a tunnel between two nodes where the traffic over the tunnel is mostly encrypted e.g. with IPsec. The problem for the protection system is that the traffic is encrypted and can be only checked on the two nodes of the tunnel where one node is mostly external. Thus, centralised protection components present in the network - e.g. the intrusion detection system at the internet gateway - cannot check the network traffic due to its encryption. 

SANA protects the network against the attacks using the encrypted VPN tunnel because it checks the network traffic with the distributed intrusion detection system from the artificial cells at least at the internal node of the VPN tunnel. When both nodes use SANA as protection system, both nodes check the traffic using the artificial cells. In addition, the node is observed so that the installation of an intrusion is detected as well as prevented. Additionally, regular checks are performed so that the nodes are checked for infections.

Our simulations show that SANA identifies the attacks, which use the VPN for attacks. In addition, the self management makes it feasible to increase the concentration of cells in these nodes because a  VPN tunnel is also a risk for the network. 

\subsection{Dynamic and Adaptive Behaviour}
The dynamic behaviour featured by the artificial cells is crucial for the performance of SANA. The artificial cells move and the self management ensures that each node is properly secured. Thus, an attacker does not always find the same configuration of the protection system and cannot be sure to use a backdoor for further attacks. The system additionally adapts to the current situation. Hotspots of attacks are identified and the concentration of artificial cells in this area is increased in order to provide more security. The collaboration between the artificial cells enables the combination of information gathered in different protection processes. SANA also adapts through analysing infections and autonomous generation of signatures in order to identify the intrusion afterwards. These points increase the performance and are not available in most common used protection components. 

\subsection{Distributed Intrusion Detection System}
\label{distrids}
The artificial cells for checking network packets implement a distributed intrusion detection system. These cells run on all nodes and evaluate the packets whether they contain an intrusion or not. The generated warnings and alerts are both sent to the administrations and processed automatically. With the right configuration, which is guaranteed through the self management, each packet is checked against all known intrusions. E.g. in all networks each packet travels at least three hops: sender, network equipment, and receiver. Thus, each hop must only check the packet against $33\%$ of all known intrusions. When the network traffic is analysed, it can be seen that in most networks the packets traverse mostly about ten hops and the configuration can be adapted to this. Also, the self management can configure the system according to the normal routing paths as well as to the routing in the network. 

The distributed intrusion detection system installed through the artificial cells works in the simulations well and identifies reliably the intrusions. Furthermore, the problem of packet loss \cite{Sch05} of current centralised approaches is solved because the dataset of known intrusions in a node is significantly reduced. 

\subsection{Security in SANA}
The distributed protection system SANA can be a risk for the network. When SANA is not properly secured, adversaries can use the system in order to run attacks, e.g. through propagating intrusions as artificial cells. Also, the artificial cells can access critical resources, which are also of special interest for an attacker. This must be prevented through a proper securing of the protection system and its components.

Section \ref{cellcomm} introduces the artificial receptors as a public private key pair describing the type and status of all components in SANA, especially from the artificial cells. These keys are extended to a distributed public key infrastructure: each resource is secured using some keys and the artificial cells have to authorise with the right keys in order to receive access. Thus, unauthorised cells cannot access the important resources. Artificial cells introduced by attackers - so called bad artificial cells - receive no access to the resources because they lack from having the right receptors. 

This approach leads to the next problem that the attackers cannot receive access to the keys of the artificial receptors. Therefore, the implementation of the security environment is important, which is introduced in section \ref{implementation} and visualised in figure \ref{pic:implementationlayer}. The system is organised in three parts: the hardware and the core parts of the operating system (operating system layer $1$), the SANA security environment with the protection components, and the operating system with the applications software. The question is, which information can be read by the adversarial when occupying the node:

\begin{itemize}
\item When an intrusion installs something in the operating system layer $1$, SANA uninstalls itself immediately from this node and quarantines the node through informing the neighbours. Thus, the intrusion has no change to receive information about the keys. Afterwards, the administrators have to check the node and bring it back to production through inserting it in the network. 
\item Most intrusion attack the Windows operating system, which is only available in the operating system part of the security environment. Using the hardware virtualisation, it is impossible to access the security environment and to access the keys. When the intrusion tries to break out, it has to change the operating system layer $1$, which is discussed in the last point. 
\item The most critical problem is an installation of an intrusion in the security environment. An arriving intrusive protection component is identified and removed. When an intrusion installs itself in the environment, integrity checks recognise this and the intrusion is identified and prevented. Due to the small size and limited features can the environment safely installed so that intrusions cannot attack it. Different techniques are e.g. formal verification and integrity checks. 
\end{itemize}

To sum up, it is hard for an adversarial to use SANA for running attacks. In addition, self checking by other protection components and especially artificial cells identify not proper working components quickly and remove them from production. 

\section{Next Steps}
There are lots of open problems and next steps in the SANA project. One important next step is to intensive the research in the cooperation between the different artificial cells. Therefore, the goal is to use the danger theory for the collaboration between different protection components in SANA. 

In the implementation, the idea of using hardware virtualisation for implementing is not enough researched. Here, the next step is to analyse the required overhead as well as to introduce the service oriented architecture in detail in order to analyse the novel features emerging out of this architecture. 

Furthermore, to extend the library of implemented protection components e.g. with honeyports is a next step in order to include more approaches of network security. With the extensions, SANA is tested against more intrusions and a testbed is built up where SANA protects a real network against real attacks. 

\section{Conclusion}
\label{secConclusion}
As it was described, the protection system SANA outperforms current protection systems. An intelligent administration and a distributed architecture with a standardised protection environment increases its performance. The distributed and dynamic framework makes it hard to attack and to break it completely down. The system is easily update and extendable where also the administration is simplified so that the system is quickly adapted to the current situation in the network. 

\section*{Acknowledgments}
SANA is currently implemented at the University of Luxembourg in the Faculty of Science, Technology, and Communication. We thank Ulrich Sorger, Zdzislaw Suchanecki, and Foued Melakessou from the University of Luxembourg for helpful discussions and the Ministre Luxembourgeois de l'education et de la recherche for additional financial support. 

\bibliographystyle{splncs}
\bibliography{paper}

\end{document}